\documentclass[12pt]{article}

\usepackage{amsmath}
\usepackage{amssymb}

\renewcommand{\section}{\secdef\sea\seb}
\newcommand{\sea}[2][default]{%
  {\vspace{0.5\baselineskip}\flushleft\bfseries #2\nopagebreak\\ \vspace{0.5\baselineskip}}}
\newcommand{\seb}[1]{\sectionmark{#1}%
  {\vspace{0.5\baselineskip}\flushleft\bfseries #1\nopagebreak\\ \vspace{0.5\baselineskip}}}

\newcommand{\bm}[1]{\boldsymbol{#1}}
\renewcommand{\Im}{\mathrm{Im}}
\renewcommand{\Re}{\mathrm{Re}}

\newcommand{\Tr}{\mathrm{Tr}}
\renewcommand{\i}{\mathrm{i}}
\newcommand{\e}{\mathrm{e}}
\renewcommand{\d}{\mathrm{d}}
\newcommand{\Scal}{\mathcal{S}}

\newcommand{\Wcal}{\mathcal{W}}
\newcommand{\Zcal}{\mathcal{Z}}
\newcommand{\Acal}{\mathcal{A}}

\newcommand{\uo}[2]{\vphantom{|}_{#1}^{\phantom{#1}#2}}
\newcommand{\oo}[1]{\vphantom{|}^{#1}_{}}
\newcommand{\uu}[1]{\vphantom{|}_{#1}^{}}
\newcommand{\pint}[1]{\int\!\mathrm{D}[#1]}
\newcommand{\fdiff}[1]{(\delta/\delta #1)}
\newcommand{\vi}{\varphi}
\newcommand{\eps}{\epsilon}
\newcommand{\mw}[1]{\langle\!\langle #1 \rangle\!\rangle}
\newcommand{\IZ}{\mathbb{Z}}

\begin{document} 
\sloppy
\thispagestyle{empty}
{\raggedright
{\large\bfseries Non-linear conductivity of charge-density-wave systems}\\
\vspace{\baselineskip}
{\bfseries C.R.~Werner and U.~Eckern}\\
\vspace{0.5\baselineskip}
Institut f\"ur Physik, Universit\"at Augsburg,
D-86135 Augsburg, Germany\\
\vspace{2\baselineskip}
\parbox[t]{154mm}{{\bfseries Abstract:} We consider the problem of sliding
motion of a charge-density-wave subject to static disorder within
an elastic medium model.
Starting with a field-theoretical formulation, which allows exact
disorder averaging, we propose a self-consistent approximation scheme
to obtain results beyond the standard large-velocity expansion.
Explicit calculations are carried out in three spatial dimensions.
For the conductivity, we find a strong-coupling regime at electrical fields
just above the pinning threshold.
Phase and velocity correlation functions scale
differently from the high-field regime, and static phase correlations
converge to the pinned-phase behaviour.
The sliding charge-density-wave is accompanied by narrow-band noise.}\\
\vspace{\baselineskip}
\vfill
published in Ann. Physik {\bfseries 6} (1997) 595-610}

\newpage
\section{1 Introduction}\noindent
The peculiar dynamics of charge-density-wave (CDW) systems
has attracted attention for more than two decades \cite{GG89}.
Among the most prominent transport properties observed
is the non-linear electric current response to a constant electric field,
the pinning of the CDW below finite threshold,
harmonic and subharmonic mode-locking in an additional oscillating drive,
and hysteretic effects.
It is well established that this behaviour has its very
origin in the interaction of the CDW with quenched
disorder in the parent substance.

The CDW state is formed through the instability of the quasi one-dimensional
electron system with respect to static density modulation below the
Peierls transition, induced by electron-phonon interaction.
It is characterized by a complex order parameter, the amplitude being
proportional to the amplitude of uniaxial electron density modulation,
and the phase modes as
low-energy collective excitations describing deformations \cite{FL78}.
Microscopic theory \cite{EG86} shows the classical character of phase dynamics,
which is usually overdamped.
Therefore, the intriguing difficulties of theoretical approaches to
CDW dynamics stem from the many degrees of freedom coupled non-linearly to
disorder.
We limit ourselves to phase dynamics throughout this paper, and
neglect a possible influence of topological defects as well.

In the following we focus on the sliding motion of the CDW driven
by a constant field.
The simplest picture describes the dynamics by a single variable.
Its equation of motion is assumed to be that of an overdamped particle
in a one-dimensional periodic potential
supplementary pulled by a constant force.
(Note the analogy to the dynamics of Josephson junctions.)
Though this model renders correctly qualitative features as the non-linear
force-velocity dependence and even pinning, the periodic potential is
an unknown input and a sinusodial potential (independent of particle velocity)
yields quantitavely incorrect results, especially for exponents.
Also it does not give any information on correlation functions.

A perturbative approach to the problem, including the spatial degrees of
freedom of the phase, was performed by Sneddon et al.~\cite{SCF82}, inspired
by earlier work on vortex lattices in type-II superconductors \cite{SH73}.
The disorder, which is considered as perturbation to the overdamped motion,
is modelled by a static random potential with vanishing mean.
In three space dimensions, the authors found in second order a correction to
Ohm's law proportional to the square root of the field.
Generalizing to arbitrary dimensions $d$, the perturbation expansion
breaks down at small fields for $d \leqslant 4$. 

The more recent view of the depinning transition as a dynamic critical
phenomenon requires a dynamic field theory,
i.e.~a formulation, where disorder averaged dynamics is mapped exactly on
a self-interacting system \cite{DD78,ER91}.
Thus the apparatus of renomalization group theory can be deployed \cite{NF92}.
Interestingly, non-standard elements remain,
e.g.~the effective periodic force correlator has to be given assertively
a cusp at its maximum.

Recent numerical \cite{SL90,MS93} and experimental \cite{MT91} work seems
to reveal a more complex scenario above threshold. 
The aim of this paper is to formulate and work out self-consistent
approximations within the dynamic formalism,
which has several advantages compared to the
standard approach \cite{LV87,BW90}.
 
In the next Section we introduce elementary notions of phase dynamics
and convenient reduced variables.
Our starting point for further investigation of non-linear dynamics
is the formulation of Eckern et al.~\cite{ER91}, which is presented
in Sec.~3.
This functional integral representation of the generating functional
combines the dynamic variables of
forward and backward time evolution in independent external fields.
Due to a peculiar normalization property, exact disorder averaging is feasible,
resulting in an effective interaction.
In Sec.~4 we prove a functional differential equation for the
generating functional of connected Green's functions, on which
self-consistent approximation schemes can be based.
It is neither restricted to a particular interaction term nor to a definite 
field-theoretical formulation.
We define especially the mean field and the
self-consistent Hartree approximation. 
The latter is used in Sec.~5 to extract the behaviour of the CDW close to
threshold.
Finally conclusions are given in Sec.~6.
In an appendix we mention a different line to systematic approximations
based on functional Legendre transforms, applied e.g.~in \cite{MS95},
and prove equivalence of a specific approximation to one of ours.

\section{2 Phase dynamics}\noindent
The CDW represents a typical low temperature state of highly anisotropic
materials.
Due to the nesting property, the electron-phonon interaction induces
correlated scattering of electrons between  two opposite parts
of the Fermi surface,
leading to both a spatial modulation of the electron density and
a lattice distortion $\propto\cos(Qx)$ with $Q=2k_F$.
The CDW direction is chosen along the $x$-axis,
and $k_F$ denotes the Fermi wave number.
Thereby a gap $\Delta$ for quasi-particle excitations is opened.
In case of incommensurability with lattice vectors, the CDW is 
free to move along the $x$-axis.
This can be described by a continous phase shift $\vi$ in
the density modulation $\propto\cos(Qx+\vi)$.
Consequently, the CDW current density is related to the phase
via $j=en_0\partial_t\vi/Q$, where $n_0$ is the condensate density.

Omitting for the moment damping effects,
the dynamics of the collective phase mode, allowing for
slow temporal and spatial variations,
is covered by the phase Hamiltonian \cite{FL78} or, alternatively, by
the action \cite{EG86,We97}
\begin{equation}
  S=n_0\int\!\d t\,\d^3r\,\Bigl\{
 \frac{\hbar m_F}{4Qv_F}\,(\partial_t\vi)^2-
 \frac{\hbar v_F}{4Q}\,(\partial_x\vi)^2-
 \frac{\hbar v_\perp}{4Q}\,(\nabla_{\!\perp}\vi)^2+
 \frac{eE_x}{Q}\,\vi\Bigr\}
 + S_U.
\end{equation}
The first part includes inertia, elastic energies, and the
electric field component $E_x$.
The condensate density calculated from microscopic theory reads
$n_0=\hbar v_FN_FQ(1-Y)$, where $v_F$, $N_F$, and $Y$ denote Fermi velocity,
density of states per spin at the Fermi surface, and the Yoshida function.
The prefactor of the time derivative is enhanced by the Fr\"ohlich mass ratio
$m_F=1+4\Delta^2/[\lambda\hbar^2\omega_Q^2(1-Y)]$.
It depends on the dimensionless electron-phonon strength $\lambda$, and the
phonon frequency $\omega_Q$.
Typically, the transverse velocity $v_\perp$ is much smaller than $v_F$.
The second part,
\begin{equation}
 S_U=n_Q\int\!\d t\,\d^3r\,\Re[\xi\exp(\i\varphi)],
\end{equation}
contains the influence of disorder, which breaks translational invariance.
The density $n_Q=2N_F\Delta/\lambda$ represents the amplitude
of the electron density modulation.
In the appropiate weak-pinning limit, disorder is simulated through 
static complex random fields $\xi$.
They can be chosen to have zero mean value, and correlations
$\langle\,\xi\,\xi\,\rangle=\langle\,\xi^\ast\xi^\ast\rangle=0$ and 
\begin{equation}
 \langle \,\xi(\bm{r})\,\xi^\ast(\bm{r}')\,\rangle =
 \frac{\hbar}{2\pi\tau_Q N_F}\,\delta(\bm{r}-\bm{r}'),
\end{equation}
where $\tau_Q$ is the characteristic time for scattering processes
between opposite parts of the Fermi surface.
Regarding damping effects,
investigations of phase dynamics can be based, motivated
phenomenologically, on the equation of motion
\begin{equation}
 n_0\gamma\,\partial_t\vi=\frac{\delta S[\vi]}{\hbar\,\delta\vi},
\end{equation}
thereby introducing the damping constant $\gamma$.

Typically, disorder leads to suppression of long-range order.
This can easily be demonstrated in the present case
by calculating the equal-time phase-phase correlations up to second order
in $\xi$. In three dimensions and specializing to the static case $E_x=0$,
they are given by 
\begin{equation}\label{iaad}
 \langle[\vi(\bm{r},0) - \vi(0,0)]^2\rangle =
 \sqrt{(x/L_0)^2 + (\bm{r}_\perp/L_\perp)^2}.
\end{equation}
The quantities
$L_0 = 4\pi^2\tau_Q \hbar N_F v_F v_\perp n_0^2/Q^2 n_Q^2$ and
$L_\perp=\sqrt{v_\perp/v_F}\,L_0$
determine the length scale, where disorder becomes relevant.
Apart from a numerical factor, this scale follows as well from minimizing
elastic and disorder energy \cite{FL78,La70}.
The linear dependence on spatial separation persists to all orders
in perturbation theory \cite{EL77}.
It has, however, become clear that perturbation theory must fail, because
the pinned CDW can access a multitude of metastable configurations.
More sophisticated approaches indicate a logarithmic growth with
distance on long scales \cite{VF84,Na90,Ko93,GL94}.

The quantities
$L_0$, $L_\perp$,
$t_0=2QL_0^2\gamma/v_F$,
$E_0=\hbar v_F/2eL_0^2$, and
$\xi_0=\hbar v_Fn_0/2QL_0^2n_Q$,
suitable for overdamped motion,
are used to introduce reduced variables for
the $x$-coordinate, transverse coordinates, time, electric field,
and random fields.
From now on we use exclusively these reduced variables,
but maintain notation. 
For later convenience some of the above equations are displayed again.
The action is then written as
\begin{equation}\label{iiaa}
  \frac{S}{\hbar} = \Lambda \int\!\d t\,\d^3r \Bigl\{
  \frac{\mu}{2}\,(\partial_t\vi)^2 - \frac{1}{2}\,
  (\nabla\vi)^2 + U(\vi) + E_x\vi \Bigr\},
\end{equation}
where
\begin{equation}\label{iiab}
 U(\vi)=\Re[\xi\exp(\i\varphi)].
\end{equation}
Correlations are given by
\begin{equation}
 \langle \,\xi(\bm{r})\,\xi^\ast(\bm{r}')\,\rangle =
 8\pi\,\delta(\bm{r}-\bm{r}'),
\end{equation}
and the equation of motion reads
\begin{equation}\label{iiac}
 \mu\,\partial_t^2\vi+\partial_t\vi - \nabla^2\vi = E_x + U'(\vi).
\end{equation}
The prefactor $\mu=m_F/2Q^2L_0^2\gamma^2$ is typically small compared to one.
Whereas $\Lambda=L_0L_\perp^2n_0\gamma$ being large, indicates a strong
suppression of thermal phase fluctuations.
For numerical values, especially concerning NbSe$_3$,
we refer to \cite{ER91,We97}.

\section{3 Generating functional}\noindent
The essential building block of dynamical field theory \cite{CY85,EP87} is the
contour along which time evolution of the system is considered:
It runs from $t=-\infty$ to $t=+\infty$, and back to $t=-\infty$.
For convenience, the contour integration is split into two
ordinary time integrations by introducing an additional index $\alpha$
denoting forward ($\alpha=1)$ and backward ($\alpha=2$) path.
The phase variables are then combined into $\vi\oo{a}=(\vi\oo{1},\vi\oo{2})$,
where $a=(\alpha,t,\bm{r})$, as well as the external fields into
$\eta\oo{a}=(\eta\oo{1},\eta\oo{2})$.
Minus signs from backward integration can be hidden in lower index
quantities, e.g.~$\eta\uu{a}=(\eta\oo{1},-\eta\oo{2})$.
This is supplemented by the convention that only indices appearing
as both upper and lower ones are subject to path summation, and 
time and space integration respectively.

Taking an average with respect to the initial state,
the complete dynamical information is included in the generating functional
\begin{equation}\label{iiiaa}
 \Zcal_U[\eta] = \pint{\vi}\,\exp
 \bigl(\i\Scal_0[\vi]+\i\Scal_{U}[\vi]+\i\eta\uu{a}\vi\oo{a}\bigr).
\end{equation}
It is obvious that the exponent essentially contains the difference
between the action (\ref{iiaa}) with variables of path one and two.
Specifically, the action of the free system is
\begin{equation}
 \Scal_0 =\frac{1}{2}\vi\oo{a}\Gamma^0_{ab}\vi\oo{b}.
\end{equation}
The free system is assumed to have been in thermal equilibrium
in the infinite past, and the disorder contribution
\begin{equation}
 \Scal_{U} = \Lambda\int\!\d t\,\d^3r\bigl\{
 U(\vi\oo{1}) - U(\vi\oo{2})\bigr\}
\end{equation}
is switched on adiabatically.
By construction, the generating functional is normalized to one for
$\eta\oo{1}=\eta\oo{2}$, independent of $\Scal_{U}$.
This most important feature can also be demonstrated
perturbatively.

A more transparent formulation is achieved by transforming to
sum and difference variables $\bar{\vi}\oo{a}$ via
$\vi\oo{b}=\bar{\vi}\oo{a}Q\uo{a}{b}$,
the matrix $Q\uo{a}{b}$ being defined through
\begin{equation}
\vi\oo{1}=\bar{\vi}\oo{2}+\bar{\vi}\oo{1}/2\Lambda,\;
\vi\oo{2}=\bar{\vi}\oo{2}-\bar{\vi}\oo{1}/2\Lambda.
\end{equation}
The inclusion of $\Lambda$ in the transformation takes into account the
smallness of fluctuations in $\vi\oo{1}-\vi\oo{2}$.
In order to keep covariance of the above expressions, we introduce
new quantities $\bar{\eta}\uu{a}=Q\uo{a}{b}\eta\uu{b}$ and
$\bar{\Gamma}^0_{ab}=Q\uo{a}{c}Q\uo{b}{d}\Gamma^0_{cd}$,
and finally suppress from now on the overlines.
Thus, using (\ref{iiab}), the disorder term reads
\begin{equation}\label{iiiab}
 \Scal_{U} = 2\Lambda\int\!\d t\,\d^3r\,
 \sin(\vi\oo{1}/2\Lambda)\,U'(\vi\oo{2}).
\end{equation}

Omitting for the moment $\Scal_{U}$ in (\ref{iiiaa}), the functional
integral is gaussian and can be calculated easily:
\begin{equation}
 \Zcal_0[\eta]=\exp[(-\i/2)\eta\uu{a}G_0^{ab}\eta\uu{b}].
\end{equation}
Both operators $\Gamma^0_{ab}$ and $G_0^{ab}$ are related by
$\Gamma^0_{ab}G_0^{bc}=\delta\uo{a}{c}$, with
$\delta\uo{a}{c}$ the identity. They have the matrix structure
\begin{equation}\label{iiiae}
 \Gamma^0_{ab}=\Biggl(\begin{array}{cc}
 \Gamma_0^K & \Gamma_0^R \\ \Gamma_0^A & 0 \end{array}\Biggr),\;
 G_0^{ab}=\Biggl(\begin{array}{cc}
 0 & G_0^A \\ G_0^R & G_0^K \end{array}\Biggr).
\end{equation}
The normalization property of the generating functional, now fulfilled for
\begin{equation}
 \eta\uu{a} = \Biggl(\begin{array}{c} E_x\\ 0\end{array}\Biggr),
\end{equation}
is a result related to the identity $G_0^{11}=0$.
This generalizes to all $n$-point functions in the interacting case.
The other entries, their Fourier transforms being displayed below,
are the retarded and advanced components
\begin{equation}
 G_0^R = (\mu\omega^2+\i\omega-\bm{q}^2)^{-1},\; G_0^A = (G_0^R)^*.
\end{equation}
Finally the Keldysh component is given by the fluctuation-dissipation
theorem (classical limit)
\begin{equation}
 G_0^K = (G_0^R-G_0^A)(\theta/\omega),
\end{equation}
where $\theta=k_BTt_0/\Lambda\hbar$.
In this formulation, the damping could be introduced easily.
Also the separation of response and correlation functions and
their possible independence in non-equilibrium is clearly visible.  
Thermal fluctuations of the phase have white noise character with amplitude
$\Gamma^K_0= 2\i\theta$.
In explicit calculations, in addition,
we have to introduce a high-frequency cut-off
in order to obtain finite local phase fluctuations.
This is physically justified by the fact that the above action
is valid only for low energy phenomena.

The phase dynamics at fixed disorder potential $U$ is characterized by the
$n$-point functions $\langle\vi\oo{a_1}..\,\vi\oo{a_n}\rangle\uu{U}$.
Given the generating functional $\Zcal_U$, they follow as $n$-time derivatives
at $\eta\uu{2}=0$, normalized to the functional itself at the same point.
Due to the normalization property $\Zcal_U[\eta\uu{2}=0]=1$,
we may readily use the disorder averaged functional
\begin{equation}
 \Zcal = \langle\Zcal_U\rangle
\end{equation}
to generate disorder averaged $n$-point functions
$\mw{\vi\oo{a_1}..\,\vi\oo{a_n}}=
\langle\,\langle\vi\oo{a_1}..\,\vi\oo{a_n}\rangle\uu{U}\rangle$.
The averaging procedure can be performed inside the functional integral.
Using the formula $\langle\exp(\i\Scal_{U})\rangle=
\exp(-\langle\Scal_{U}\Scal_{U}\rangle/2)=\exp(\i\Scal_{1})$,
where the last equation defines the effective interaction
\begin{equation}\label{iiiac}
  \Scal_{1} = 2\pi\i\int\!\d^3r\d\tau\d\bar{\tau}\;(2\Lambda)^2
  \sin\bigl(\vi\oo{1}/2\Lambda\bigr)
  \sin\bigl(\vi\oo{\bar{1}}/2\Lambda\bigr)
  \cos\bigl(\vi\oo{2}-\vi\oo{\bar{2}}\bigr),
\end{equation}
we finally obtain \cite{ER91}
\begin{equation}\label{iiiad}
  \Zcal[\eta] = \pint{\vi}\,\exp
  \bigl(\i\Scal_0[\vi]+\i\Scal_1[\vi]+\i\eta\uu{b}\vi\oo{b}\bigr).
\end{equation}
We remark that the field-theoretic approach to
classical dynamics \cite{MSR73,Ph77}, starting from (\ref{iiac}), leads to
the contributions (\ref{iiiab}) and (\ref{iiiac})
in the limit $\Lambda\to\infty$.

\section{4 Self-consistent approximations}\noindent
In the last chapter we derived an effective action that governs
the disorder averaged phase dynamics.
However it contains the non-linear term (\ref{iiiac}), and
therefore one has to resort to an approximative analysis of the problem.
The first step in our strategy is to write (\ref{iiiac}) in the form 
\begin{equation}\label{ivaa}
 \Scal_{1} = -\i\,\Tr^A\exp\bigl(\i\zeta^A_a\vi\oo{a}\bigr).
\end{equation}
Here the explicit structure is encoded in the symbols
\begin{align}
 \Tr^A &= -\pi\int\!\d^3r_A\d\tau_A\d\bar{\tau}_A\!\!\!
 \sum_{\eps_A,\bar{\eps}_A=\pm 1/2\Lambda}
 \!\!\!(2\eps_A\bar{\eps}_A)^{-1}, \\
 \zeta^A_a &= \delta(\bm{r}_a-\bm{r}_A)\,
 \Biggl[\,\delta(t_a-\tau_A)\,
 \Biggl(\begin{array}{c} \eps_A\\ 1\end{array}\Biggr) -
 \delta(t_a-\bar{\tau}_A)\,
 \Biggl(\begin{array}{c} \bar{\eps}_A\\1\end{array}\Biggr)\,\Biggr],
\end{align}
where $A=(\bm{r}_A,\tau_A,\bar{\tau}_A,\eps_A,\bar{\eps}_A)$
combines internal variables being summed and integrated respectively.
Then recalling that the integral of a derivative vanishes identically,
we perform the derivative in
$\pint{\vi}\,\fdiff{\vi}\uu{a}\exp(\i\Scal[\vi]+\i\eta\uu{b}\vi\oo{b})=0$
to get
\begin{equation}
 \pint{\vi}\,\bigl(\fdiff{\vi}\uu{a}\Scal[\vi]+\eta\uu{a}\bigr)
 \exp\bigl(\i\Scal[\vi]+\i\eta\uu{b}\vi\oo{b}\bigr)=0.
\end{equation}
A derivative symbol with lower index is defined as derivative with respect
to the specified upper index quantity and vice versa.
Representation (\ref{ivaa}) of the interaction term leads immediately to
\begin{equation}
 -\i\Gamma^0_{ab}\fdiff{\eta}\oo{b}\Zcal[\eta]+
 \Tr^A\zeta^A_a\Zcal[\eta+\zeta^A]+\eta\uu{a}\Zcal[\eta]=0.
\end{equation}
Finally introducing the generating functional $\Wcal$
of connected Green's functions, defined through $\Zcal=\exp(\i\Wcal)$,
we arrive at
\begin{equation}\label{ivab}
 \fdiff{\eta}\oo{a}\Wcal[\eta] = -G_0^{ab}\eta\uu{b}
 -G_0^{ab}\,\Tr^A\zeta^A_b
 \exp\bigl(\i\delta\Wcal^A[\eta]\bigr),
\end{equation}
where
\begin{equation}
 \delta\Wcal^A[\eta] = \Wcal[\eta+\zeta^A]-\Wcal[\eta].
\end{equation}
This is our central result, on which approximation schemes are based.
Equation (\ref{ivab}) determines directly the response to
external fields. Further derivatives allow access to linear response
and correlation functions, and so on.
A straightforward procedure to obtain systematic approximations is to
truncate the expansion
\begin{equation}\label{ivac}
 \delta\Wcal^A[\eta]=
 \sum_{n=1}^{\infty}\frac{1}{n!}\,\zeta^A_{a_1}..\,\zeta^A_{a_n}\,
 \fdiff{\eta}\oo{a_1}..\,\fdiff{\eta}\oo{a_n}\Wcal[\eta].
\end{equation}

The lowest order approximation,
from now on referred to as mean-field approximation,
keeps only the first term ($n=1$) in (\ref{ivac}).
The quantities $\phi\oo{a}=\fdiff{\eta}\oo{a}\mathcal{W}$
and $G\oo{ab}=-\fdiff{\eta}\oo{a}\fdiff{\eta}\oo{b}\mathcal{W}$,
defined for $\eta\uu{a}=(E_x,0)$, are determined by (\ref{ivab}) and its
derivative:
\begin{align}
 \phi\oo{a} &= -G_0^{ab}\eta\uu{b}
 -G_0^{ab}\bigl[\Tr^A\zeta^A_b\exp\bigl(\i\zeta^A_c\phi\oo{c}\bigr)\bigr],\\
 G\oo{ab} &= G_0^{ab}+G_0^{ac}
 \bigl[-\i\,\Tr^A\zeta^A_c\zeta^A_d\exp\bigl(\i\zeta^A_e\phi\oo{e}\bigr)\bigr]
 G\oo{db}.
\end{align}
Straightforward calculation shows that the first equation is simply
the saddle-point condition.
In the second one the self-energy $\Sigma\uu{cd}$, which equals
the expression in squared brackets, is merely dependent on the mean value
$\phi\oo{a}$, and therefore neglects fluctuation effects as well.
We add here that the self-energy follows from
Dyson equation $\Gamma\uu{ab}G\oo{bc}=\delta\uo{a}{c}$
through $\Gamma\uu{ab}=\Gamma^0_{ab}-\Sigma\uu{ab}$.

The second order approximation keeps the first two terms
($n=1$ and $n=2$) in (\ref{ivac}).
Unfortunately the analogous equations do not form a closed system.
The third derivative of $\Wcal$ couples to the self-energy.
Omitting this, we obtain the so called self-consistent Hartree appoximation:
\begin{align}\label{ivad}
 \phi\oo{a} &= -G_0^{ab}\eta\uu{b}
 -G_0^{ab}\bigl[\Tr^A\zeta^A_b\exp\bigl(\i\zeta^A_c\phi\oo{c}
 -(\i/2)\zeta^A_cG\oo{cd}\zeta^A_d\bigr)\bigr],\\ \label{ivae}
 G\oo{ab} &= G_0^{ab}+G_0^{ae}
 \bigl[-\i\,\Tr^A\zeta^A_e \zeta^A_f
 \exp\bigl(\i\zeta^A_c\phi\oo{c}
 -(\i/2)\zeta^A_cG\oo{cd}\zeta^A_d\bigr)\bigr]
 G\oo{fb}.
\end{align}
Again the expression in squared brackets in the second equation
is identified as the self-energy.
In contrast to the mean-field equations, the above approximation takes into
account gaussian fluctuations in a self-consistent manner.

\section{5 Results and discussion}\noindent
We focus now on the case, where the CDW
is moving uniformly with velocity $\bar{v}$ in
a constant field $E_x$, i.e.~$\mw{\vi}=\bar{v}t$.
This translates into $\eta\uu{a}=(E_x,0)$ and
$\phi\oo{a}=(0,\mw{\vi})$ for the representation
chosen in the preceding section.
Furthermore, $G\oo{ab}$, which solely depends on differences of 
time and space arguments, has the structure (\ref{iiiae}).

As it turns out that mean-field equations give no corrections to free motion,
except for correlations,
we consider now the self-consistent Hartree approximation.
Inspection of these equations shows immediately that the parameter $\Lambda$
merely appears as a non-relevant short-time cut-off.
Therefore we display equations (\ref{ivad},\ref{ivae})
in the classical limit $\Lambda\to\infty$:
\begin{equation}\label{vaa}
 \bar{v} = E_x +
 4\pi\int\!\d\tau\;\e^{-M(\tau)}\,\sin(\bar{v}\tau)\;G^R(0,\tau).
\end{equation}
The self-energies are
\begin{align}\label{vab}
 \Sigma^R(\bm{r},t) &= 4\pi\,\delta(\bm{r})\,
 \bigl[\,V(t)\,G^R(0,t) - \delta(t)\int\!\d\tau\,V(\tau)\,
 G^R(0,\tau)\bigr],\\
 \Sigma^K(\bm{r},t) &= -4\pi\i\,\delta(\bm{r})\,V(t),  
\end{align}
where
\begin{align}
 M(t) &= \i[G^K(0,0)-G^K(0,t)],\\ \label{vad}
 V(t) &= \e^{-M(t)}\,\cos(\bar{v}t).
\end{align}
The quantity $M(t)$ equals the correlation function
$(1/2)\,\mw{[\delta\vi(0,t)-\delta\vi(0,0)]^2}$,
where $\delta\vi=\vi-\mw{\vi}$.
We note that the selfenergies are only frequency dependent, so
wave-vector integration can be executed in Fourier representations
of $G^R$ and $G^K$.
After some algebra, the Fourier transform of (\ref{vab}) reads
\begin{equation}\label{vaz}
 \Sigma^R_\omega = \int\!\frac{\d\nu}{2\pi}
 \bigl(\Sigma^K_{\nu-\omega}-\Sigma^K_\nu\bigr) \frac{Z^R_\nu}{4\pi},
\end{equation}
where
\begin{equation}\label{vazz}
  Z^R_\nu=\sqrt[+]{\mu\nu^2+\i\nu-\Sigma^R_\nu}
\end{equation}
is specific to $d=3$.
By convention, $\Im\sqrt[+]{\vphantom{x}}\geq 0$,
i.e.~the cut is on the positive real axis.
In addition we find
\begin{equation}\label{vac}
 M(t) = \int\!\frac{\d\nu}{2\pi}\bigl(1-\e^{-\i\nu t}\bigr)
 \bigl(-\i\Gamma^K_\nu\bigr)
 \bigl(8\pi\,\Im\, Z^R_\nu\bigr)^{-1};
\end{equation}
recall that $\Gamma\oo{K}=\Gamma^K_0-\Sigma\oo{K}$.
Thus the two self-energy components are interwoven in a non-linear fashion.

From now on we neglect the temperature, i.e. we set $\Gamma^K_0=0$.
In doing so we do not miss any qualitative features,
because long-range order is not destructed by thermal noise for $d>2$.
Numerical simplifications in this case stem from the fact
that $M(t)$ and $V(t)$ are now periodic functions in time, without any
superposed short time transients.
This is reflected by the ansatz
$\Sigma^K_\omega = -\i(2\pi)^2\sum_nV_n\,\delta(\omega-n\bar{v})$, where
summation runs over all integers.
Note that the $n=0$ contribution induces time persistent correlations.
The numbers
$V_n=(\bar{v}/\pi)\int_{-\pi/\bar{v}}^{\pi/\bar{v}}\!\d t\, V(t)\,
\e^{\i n\bar{v}t}$ are the Fourier coefficents of $V(t)$ and
determine self-consistently $M(t)=\sum_{n\neq 0}V_n
(1-\e^{\i n\bar{v}t})(4\,\Im\, Z^R_{n\bar{v}})^{-1}$.
It turns out that quantities at frequencies $n\bar{v}$, $n\in\IZ$, are
sufficient to form a closed set of equations.
In this case Eq.~(\ref{vaz}) reads
$\Sigma^R_{n\bar{v}} = (1/2\i)
\sum_m V_m( Z^R_{m\bar{v}+n\bar{v}}- Z^R_{m\bar{v}})$.
Given $\bar{v}$, the values of the $V_n$, $\Sigma^R_{n\bar{v}}$ or
$Z^R_{n\bar{v}}$ respectively, are calculated by numerical iteration.
Then the force-velocity dependence (\ref{vaa}) follows from
\begin{equation}
 \bar{v}+(1/2)\sum_nV_n\,\Im\, Z^R_{n\bar{v}}=E_x.
\end{equation}
Due to the square-root in the expression (\ref{vazz}), a small
inertia $\mu$ is irrelevant at not too high velocities.
That case is easily treated by neglecting the self-energies in a first
iteration of the self-consistent equations. 
For our purposes the limit $\mu\to 0$ is appropriate and used further on.

The result is shown in Figs.~\ref{fig1} and \ref{fig2}, where
velocities $\bar{v}$ down to $10^{-5}$ are considered.
The non-linear conductivity $\bar{v}/E_x$ develops an upward curvature
below $E_x\approx 0.5$ or $\bar{v}\approx 0.1$, respectively. As shown
by subsequent numerical results,
this marks the crossover between the high- and low-velocity regime.
A sharp drop in velocity appears at $E_x\approx 0.25$ in Fig.~\ref{fig2}.

Further information is given by the equal-time phase correlation function
\begin{equation}
 2M(\bm{r})=\mw{[\delta\vi(\bm{r},0)-\delta\vi(0,0)]^2}.
\end{equation}
It is connected
to the Green's function $G^K$ via $M(\bm{r})=\i[G^K(0,0)-G^K(\bm{r},0)]$.
The long-range behaviour, $2M(\bm{r})\simeq|\bm{r}|/L_1$, is determined
by the length $L_1=2/V_0$.
Hence phase coherence is absent in the driven CDW as well.
In the high-velocity regime, where $L_1\simeq(8\bar{v})^{1/2}$, this
spatial dependence is valid except for short distances $|\bm{r}|\lesssim 1$.
Figure \ref{fig3} shows $L_1$ as a function of velocity in the considered range,
where $L_1$ is typically one order of magnitude larger than the bare $L_0$.
It increases again at smaller velocities in a power law fashion,
$L_1\propto\bar{v}^{-0.12}$.

The spatial dependence of phase correlations for small velocities is depicted
in Fig.~\ref{fig4}.
Clearly three regimes as a function of distance appear:
\begin{equation}
 2M(\bm{r})\simeq
 \begin{cases}
 |\bm{r}| & \text{for $|\bm{r}|\lesssim 1$,}\\
 2\ln|\bm{r}| & \text{for $1\lesssim |\bm{r}|\lesssim L_1$,}\\
 |\bm{r}|/L_1 & \text{for $L_1\lesssim |\bm{r}|$.}
 \end{cases}
\end{equation}
The prefactor two of the logarithm, here determined within an error smaller
than one percent, is identical to that in the static CDW
as calculated in \cite{Ko93,GL94} by the replica method.

At high velocities the asymptotic behaviour of the normalized equal-time
velocity correlation function,
\begin{equation}
 C(\bm{r})=\mw{\delta\dot{\vi}(\bm{r},0)\delta\dot{\vi}(0,0)}/\bar{v}^2,
\end{equation}
is given by $C(\bm{r})\simeq L_2\exp(-\tilde{r})\sin(\tilde{r})/\tilde{r}$,
if expressed as a function of the rescaled distance $\tilde{r}=|\bm{r}|/L_2$.
The new length $L_2\simeq(\bar{v}/2)^{-1/2}$, determines the decay of the
oscillating velocity correlations.
Thus, in contrast to the phase itself, velocity coherence is maintained
in the sliding CDW.

Inspection of $C(\bm{r})$ in Fig.~\ref{fig5} shows that the situation is more
complex in the low-velocity regime.
Diverging relative on-site fluctuations with an approximate
power law $C(0)\propto\bar{v}^{-0.75}$ are found.
In the following, we define a length $L_2$ as in the high-velocity regime. 
On shorter distances,
a velocity independent exponential attenuation is exhibited,
which changes on larger distances into an exponential-like decay,
superposed by oscillations.
The long-range part of $C(\bm{r})$ at different velocities can be projected
on a single curve by rescaling distance through $\tilde{r}=|\bm{r}|/L_2$.
Surprisingly, $L_2$ exhibits a similar power law as $L_1$ as a function of
$\bar{v}$, although the prefactor is somewhat smaller.  
A rather good choice for the overall functional form is
\begin{equation}
 C(\bm{r})\simeq
 \begin{cases}
 C(0)\exp(-|\bm{r}|/2) & \text{for $|\bm{r}|\lesssim L_2$,}\\
 \exp(-\tilde{r})\sin(\tilde{r})/\tilde{r} &
 \text{for $L_2\lesssim |\bm{r}|$.}
 \end{cases}
\end{equation}

Our length $L_2$ corresponds to $\xi$ as introduced in Eq.~(1.5) in \cite{NF92}.
From the static distortions of the phase (1.20) in \cite{NF92}, we read off
the correspondence of our $L_1$ to $\xi^{\eta_s}$,
where the exponent $\eta_s=0$ is given in (1.21). 
This is in contradiction to our scenario, which would give $\eta_s=1$,
and also to known results for the pinned CDW.

The frequency spectrum of velocity correlation shows
narrow-band-noise due to time persistent correlations:
The zero wave-vector component of the Fourier transformed
$C(\bm{r},t)=\mw{\delta\dot{\vi}(\bm{r},t)\delta\dot{\vi}(0,0)}/\bar{v}^2$,
determines the noise spectrum
$S(\omega)=\int\d^3r\d t\,C(\bm{r},t)\exp(\i\omega t)$.
We obtain
$S(\omega)=\i(\omega/\bar{v})^2G^R_{0,\omega}\Sigma^K_\omega G^A_{0,\omega}$.
This expression contains only peaks at integer multiples of $\bar{v}$,
i.e. $S(\omega)=(2\pi)^2\sum_{n\neq 0}S_n\,\delta(\omega-n\bar{v})$,
reflecting the periodicity of the sliding state.
Its first harmonic $S_1$ is $\propto\bar{v}^{-2}$ at high, and approximately
$\propto\bar{v}^{-0.37}$ at low velocities.

Now we relate the quantities characterizing the velocity
correlations to the velocity coherence length $L_2$.
Thus $C(0)\propto L_2$ at high, and $\propto L_2^6$ at low velocities.
Correspondingly, $S_1\propto L_2^4$ at high,
and $\propto L_2^3$ at low velocities, which is in agreement with 
the behaviour of dynamic correlations (1.18) in \cite{NF92}.

In closing this section, we discuss the time dependence of force correlations.
The interaction term (\ref{iiiac}) contains the
fluctuation component $\vi^1$ quadratically (for $\Lambda\to\infty$)
and the force-force factor $\cos(\vi^2-\vi^{\bar{2}})$
in which the center-of-mass component $\vi^2$ appears.
It follows that its mean value is proportional to the second derivative
of the vertex functional with respect to $\vi^1$ or the self-energy
component $\Sigma_{1\bar{1}}$ respectively.
Actually, $V(t)=\mw{\cos[\vi(0,t)-\vi(0,0)]}$
coincides with the definition (\ref{vad}).
Its renormalization from the high-velocity limit $V(t)\simeq\cos(\bar{v}t)$
is shown in Fig.~\ref{fig6}, where the peak at $t=0$ sharpens considerably
at low velocities.
Except for this singular behaviour, the primary dependence at low velocities
appears to be a scaling of the global amplitude $\propto\bar{v}^{0.14}$.
Whether the width of the peak vanishes on further lowering $\bar{v}$
has to remain open.

\section{6 Conclusions}\noindent
We have considered the sliding CDW in presence of quenched disorder within
phase dynamics.
Our starting point is a field theoretical technique, which allows disorder
averaging directly in the generating functional, thereby introducing an
effective interaction.
To obtain explicit results, we have developed a
self-consistent approximation scheme.
In three space dimensions, the sliding regime at average velocity $\bar{v}$
is characterized by linearly growing phase fluctuations with spatial
distance. The scale is set by the diverging phase
correlation length $L_1\propto\bar{v}^{1/2}$ for high, and
$L_1\propto\bar{v}^{-0.12}$ for low velocities.
For low velocities a range with logarithmic increase with distance develops,
as known from the pinned CDW.
Nevertheless, velocity coherence is maintained, and the sliding CDW
is accompanied by narrow-band-noise.
Velocity correlations show oscillating decay with distance under an
exponential-like envelope.
The corresponding length $L_2$ is $\propto\bar{v}^{-1/2}$
for high, and $\propto\bar{v}^{-0.12}$ for low velocities.
The high-velocity behaviour of correlations is known from perturbation theory.
For low velocities the exponents of $L_1$ and $L_2$ coincide within few percent.
In addition the above behaviour of the velocity correlation
is only the asymptotics for long distances $\gtrsim L_2$, whereas for distances
$\lesssim L_2$ an exponential attenuation on a velocity independent
characteristic length appears.
The amplitude of the first harmonic of the narrow-band-noise normalized to
the mean velocity is proportional to the coherence volume $L_2^3$.

Our approximation has identified a new regime at low velocities with
different scaling properties, although it is possibly not applicable
in the critical regime close to threshold.
Direct comparison with experiment is difficult because of the possibility of
phase slip and finite size effects in real materials, but
numerical simulations appear to give a similar picture.
Finally we represented a distinct approach to systematic approximations,
and proved equivalence to our approximation scheme.

{\bf Acknowledgements:}
We thank J.~M\"ullers for several helpful discussions,
and P.~Schmitteckert for computational advice.

\newpage
\section{Appendix}\noindent
In this paper the pivotal element for generating systematic approximation
is a functional differential equation for $\Wcal$.
Evaluation proceeds in assuming that interaction induced
fluctuations $\delta\Wcal^A$ are restricted to a finite order 
Taylor expansion.
First order expansion leads straightforwardly to a mean-field description.
Whereas to second order an additional approximation has to be executed,
namely omitting third order derivatives of $\Wcal$, to close the equations.

Recently M\"ullers und Schmid \cite{MS95} applied functional Legendre
transform techniques to improve analysis of vortex dynamics in disordered
type-II superconductors.
Therein a Legendre transform is performed for both
an external field $\eta\uu{a}$ and a quadratic source $K\uu{ab}$.
Of course the essential approximation lies in the fact that
the action functional is calculated to first order perturbation theory.
In the following we show equivalence of the above to our
self-consistent Hartree approximation.
Starting from the generating functional
\begin{equation}\label{viaa}
 \Wcal[\eta,K]=-\i\ln\pint{\vi}\,\exp\bigl(\i\Scal[\vi]
 +\i\eta\uu{a}\vi\oo{a}+(\i/2)\vi\oo{a}K\uu{ab}\vi\oo{b}\bigr),
\end{equation}
new variables $\phi\oo{a}$ and $G\oo{ab}$ are defined via
\begin{equation}\label{viab}
 \fdiff{\eta}\oo{a}\Wcal = \phi\oo{a},\;
 \fdiff{K}\oo{ab}\Wcal = (1/2)(\i G\oo{ab}+\phi\oo{a}\phi\oo{b}).
\end{equation}
At the end of the calculation one sets $K\uu{ab}=0$ to extract physical
information.
The desired action functional
\begin{equation}
 \Acal[\phi,G] = \Wcal[\eta,K] - \eta\uu{a}\phi\oo{a} 
 -(1/2)\phi\oo{a}K\uu{ab}\phi\oo{b} - (1/2)K\uu{ab}G\oo{ba}
\end{equation}
is the Legendre transform of (\ref{viaa}),
where $\eta\uu{a}$ and $K\uu{ab}$ are to be eliminated using (\ref{viab}). 
The derivatives obey the inverse relations
\begin{equation}\label{viac}
 \fdiff{\phi}\uu{a}\Acal = -\eta\uu{a}-K\uu{ab}\phi\oo{b},\; 
 \fdiff{G}\uu{ab}\Acal = -(\i/2)K\uu{ab}.
\end{equation}
In order to proceed the action $\Scal[\phi]=\Scal_0[\phi]+\Scal_1[\phi]$
is now composed of a free part 
$\Scal_0=(1/2)\phi\oo{a}\Gamma^0_{ab}\phi\oo{b}$
and an interaction $\Scal_1=-\i\,\Tr^A\exp(\i \zeta^A_a\vi\oo{a})$.
It has been shown that 
\begin{gather}\nonumber
 \Acal[\phi,G] = \Scal[\phi] 
 - (\i/2)\delta\uo{a}{b}\ln \Gamma^0_{bc}G\oo{ca}
 - (\i/2)\bigl(\delta\uo{a}{a}-
 G\oo{ab}\fdiff{\phi}\uu{b}\fdiff{\phi}\uu{a}\Scal[\phi]\,\bigr)\\
 \label{viad}
 -\,\i\ln\pint{\vi}\exp\bigl((\i/2)\vi\oo{a}\Gamma\uu{ab}\vi\oo{b}
 +\i\Scal_2[\phi,\vi]\,\bigr)\Big|_\mathrm{2PI},
\end{gather}
where the functional integral is normalized to its free part, and the
new interaction term $\Scal_2[\phi,\vi]$ equals $\Scal_1[\phi+\vi]$ minus
its Taylor expansion up to second order in $\vi^a$.
The index 2PI restricts a diagrammatic perturbation theory
to two particle irreducible contributions.
As usual, $\Gamma\uu{ab}G\oo{bc}=\delta\uo{a}{c}$.
Finally expanding the logarithm in (\ref{viad}) up to first order in
$\Scal_2$, the following result is found:
\begin{gather}\nonumber
 \Acal[\phi,G] = (1/2)\phi\oo{a}\Gamma^0_{ab}\phi\oo{b} 
 - (\i/2)\delta\uo{a}{b}\ln \Gamma^0_{bc}G\oo{ca}
 - (\i/2)(\delta\uo{a}{a}-\Gamma^0_{ab}G\oo{ba}) \\
 -\,\i\,\Tr^A\exp\bigl(\i\zeta^A_a\phi\oo{a}
 -(\i/2)\zeta^A_aG\oo{ab}\zeta^A_b\bigr).
\end{gather}
Performing the derivatives (\ref{viac}) of this quantity leads to the
conditions (\ref{ivad}) and (\ref{ivae}).

To compare the two approaches, one may argue that 
both determine minimal self-consistent approximations including fluctuations. 
Whereas the above Legendre transform is truncated at first order
in the interaction, our method restricts explicitly interaction induced
fluctuations to gaussian ones.

\newpage
\setlength{\unitlength}{0.1bp}
\begin{picture}(3600,2160)(0,0)
\put(2008,1){\makebox(0,0){$E_x$}}
\put(100,1180){%
\makebox(0,0)[b]{\shortstack{$\bar{v}/E_x$}}%
}
\put(1950,600){\makebox(0,0){c}}
\put(1600,800){\makebox(0,0){a}}
\put(1250,1000){\makebox(0,0){b}}
\put(3417,151){\makebox(0,0){10}}
\put(2009,151){\makebox(0,0){1}}
\put(600,151){\makebox(0,0){0.1}}
\put(540,2109){\makebox(0,0)[r]{1}}
\put(540,356){\makebox(0,0)[r]{0}}
\end{picture}
\vspace{\baselineskip}\\
Fig.~\ref{fig1}:\refstepcounter{figure}\label{fig1}
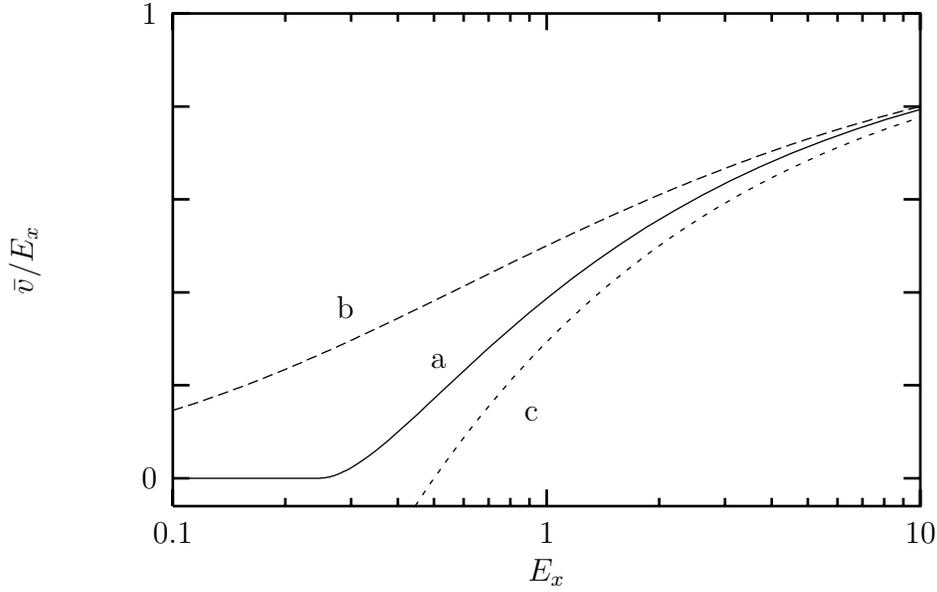
Non-linear conductivity $\bar{v}/E_x$ vs.~electric field $E_x$
in (a) self-consistent Hartree approximation, (b) Hartree approximation using
the free Green's function $G^{ab}_0$, and (c) first order perturbation theory.

\vspace{2\baselineskip}
\setlength{\unitlength}{0.1bp}
\begin{picture}(3600,2160)(0,0)
\put(2008,1){\makebox(0,0){$E_x$}}
\put(100,1180){%
\makebox(0,0)[b]{\shortstack{$\bar{v}$}}%
}
\put(3417,151){\makebox(0,0){0.5}}
\put(2478,151){\makebox(0,0){0.4}}
\put(1539,151){\makebox(0,0){0.3}}
\put(600,151){\makebox(0,0){0.2}}
\put(540,2109){\makebox(0,0)[r]{$10^{-1}$}}
\put(540,1645){\makebox(0,0)[r]{$10^{-2}$}}
\put(540,1180){\makebox(0,0)[r]{$10^{-3}$}}
\put(540,716){\makebox(0,0)[r]{$10^{-4}$}}
\end{picture}
\vspace{\baselineskip}\\
Fig.~\ref{fig2}:\refstepcounter{figure}\label{fig2}
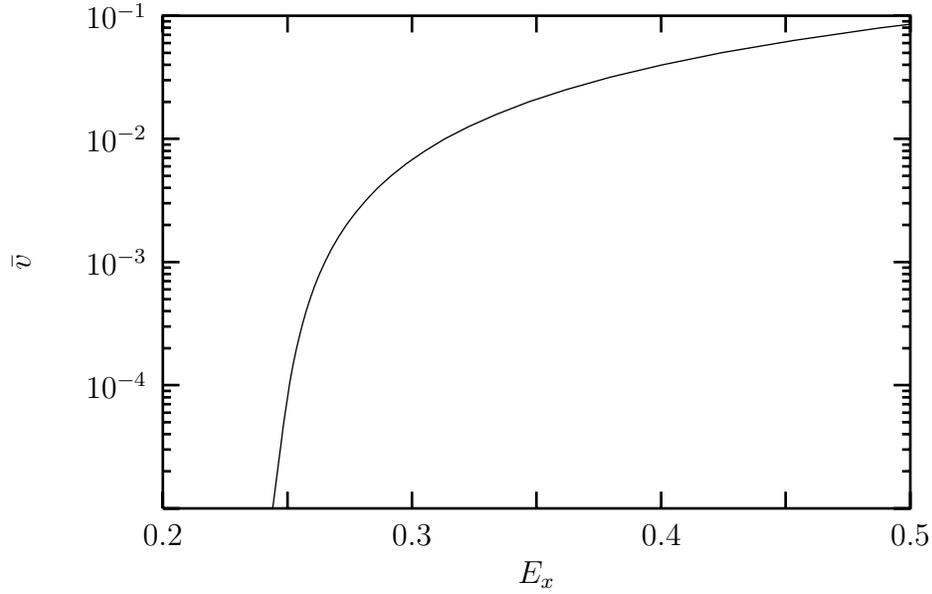
Average velocity $\bar{v}$ vs.~electric field $E_x$ in self-consistent
Hartree approximation. Note the logarithmic scale on the vertical axis.

\newpage
\setlength{\unitlength}{0.1bp}
\begin{picture}(3600,2160)(0,0)
\put(2008,1){\makebox(0,0){$\bar{v}$}}
\put(100,1180){%
\makebox(0,0)[b]{\shortstack{$L_1$}}%
}
\put(2948,151){\makebox(0,0){$10^0$}}
\put(2009,151){\makebox(0,0){$10^{-2}$}}
\put(1070,151){\makebox(0,0){$10^{-4}$}}
\put(540,2109){\makebox(0,0)[r]{30}}
\put(540,1223){\makebox(0,0)[r]{10}}
\put(540,251){\makebox(0,0)[r]{3}}
\end{picture}
\vspace{\baselineskip}\\
Fig.~\ref{fig3}:\refstepcounter{figure}\label{fig3}
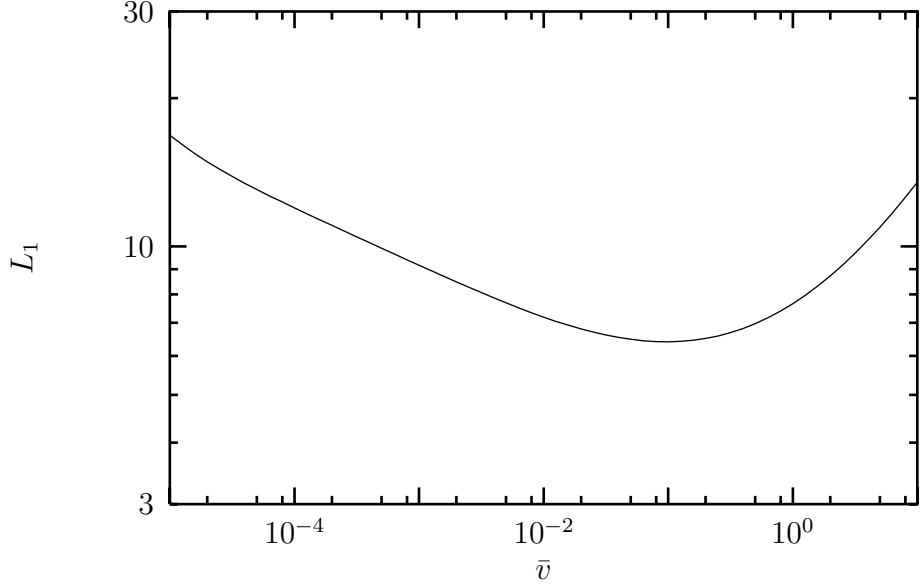
Correlation length $L_1$ vs.~average velocity $\bar{v}$ in self-consistent
Hartree approximation. Asymptotic power laws are $L_1\propto\bar{v}^{-0.12}$
for low, and $L_1\propto\bar{v}^{0.5}$ for high velocities.

\vspace{2\baselineskip}
\setlength{\unitlength}{0.1bp}
\begin{picture}(3600,2160)(0,0)
\put(1054,1506){\makebox(0,0)[r]{$10^{-1}$}}
\put(1054,1616){\makebox(0,0)[r]{$10^{-2}$}}
\put(1054,1726){\makebox(0,0)[r]{$10^{-3}$}}
\put(1054,1836){\makebox(0,0)[r]{$10^{-4}$}}
\put(1054,1946){\makebox(0,0)[r]{$10^{-5}$}}
\put(2008,1){\makebox(0,0){$|\bm{r}|$}}
\put(100,1180){%
\makebox(0,0)[b]{\shortstack{$2M(\bm{r})$}}%
}
\put(3417,151){\makebox(0,0){100}}
\put(2009,151){\makebox(0,0){10}}
\put(600,151){\makebox(0,0){1}}
\put(540,2109){\makebox(0,0)[r]{10}}
\put(540,251){\makebox(0,0)[r]{0}}
\end{picture}
\vspace{\baselineskip}\\
Fig.~\ref{fig4}:\refstepcounter{figure}\label{fig4}
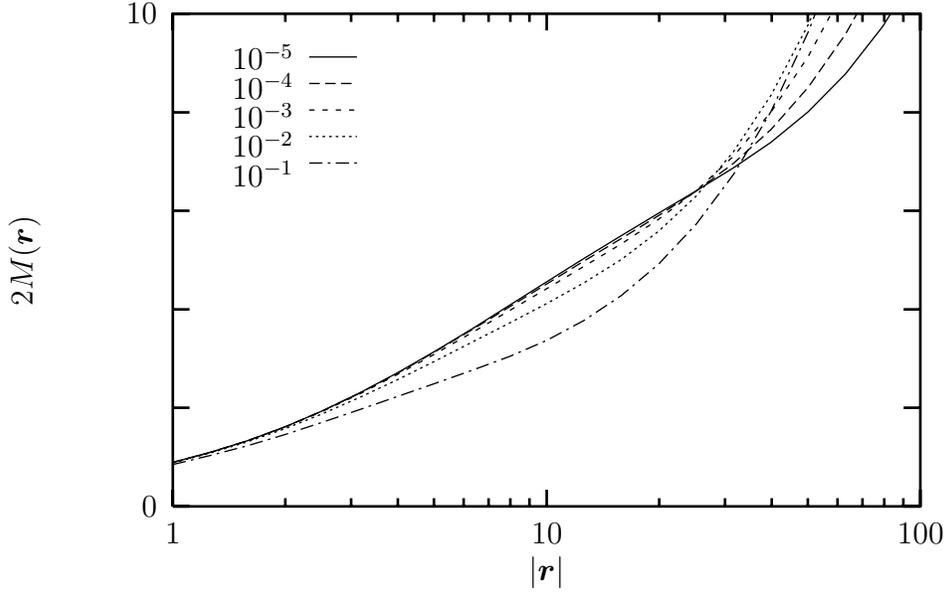
Static phase correlation
$2M(\bm{r})=\mw{[\delta\vi(\bm{r},0)-\delta\vi(0,0)]^2}$ vs.~distance
$|\bm{r}|$ in self-consistent Hartree approximation.
The velocities $\bar{v}$ are displayed in the inset.
The middle part equals asymptotically $2\ln|\bm{r}|$.

\newpage
\input{figs/fig5.tex}\vspace{\baselineskip}\\
Fig.~\ref{fig5}:\refstepcounter{figure}\label{fig5}
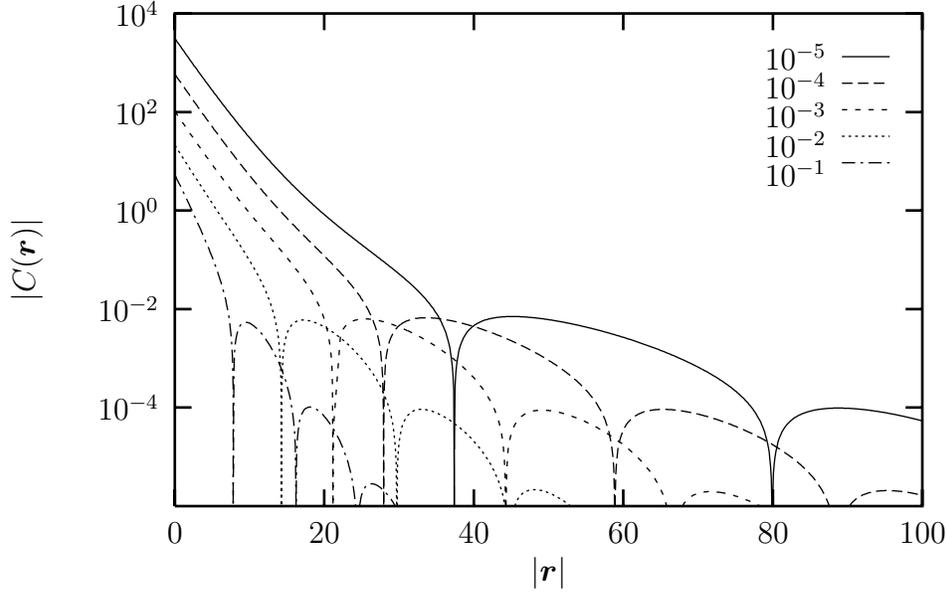
Absolute value of the static velocity correlation
$C(\bm{r})=\mw{\delta\dot{\vi}(\bm{r},0)\delta\dot{\vi}(0,0)}/\bar{v}^2$
vs.~distance $|\bm{r}|$ in self-consistent Hartree approximation.
The velocities $\bar{v}$ are displayed in the inset.
Sharp drops indicate a change of sign.

\vspace{2\baselineskip}
\input{figs/fig6.tex}\vspace{\baselineskip}\\
Fig.~\ref{fig6}:\refstepcounter{figure}\label{fig6}
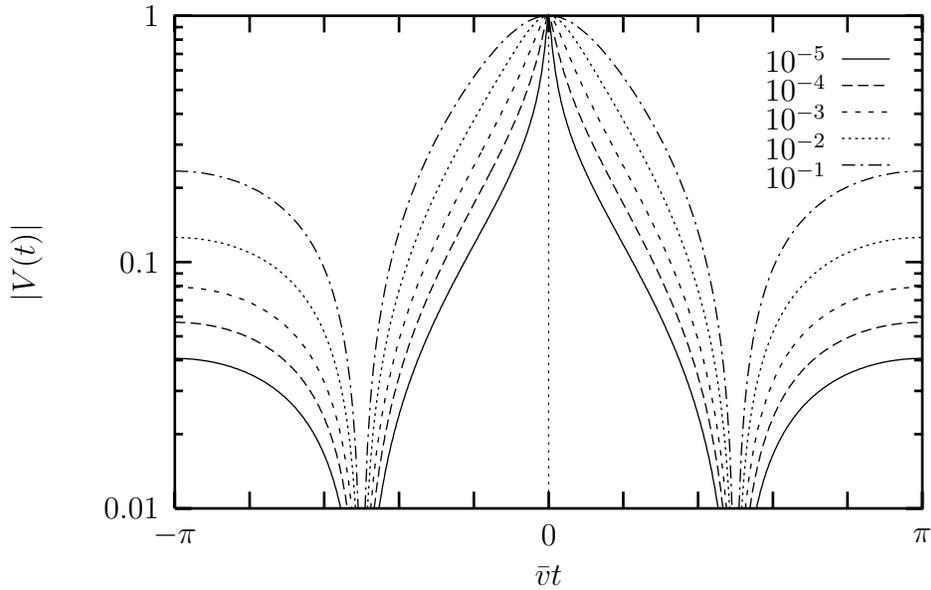
Absolute value of the renormalized force correlation $V(t)$ vs.~time over
one period $2\pi/\bar{v}$ in self-consistent Hartree approximation.
The high velocity limit is $V(t)\simeq\cos(\bar{v}t)$.
The velocities $\bar{v}$ are displayed in the inset.
Sharp drops indicate a change of sign.
\end{document}